# Coercive Field and Magnetization Deficit in $Ga_{1-x}Mn_xAs$ Epilayers

S. J. Potashnik, K. C. Ku, S. H. Chun, R. F. Wang, M. B. Stone, N. Samarth,

P. Schiffer[*]

Department of Physics and Materials Research Institute, Pennsylvania State University,

University Park, PA 16802

**Abstract**

We have studied the field dependence of the magnetization in epilayers of the diluted magnetic semiconductor $Ga_{1-x}Mn_xAs$ for $0.0135 < x < 0.083$. Measurements of the low temperature magnetization in fields up to 3 T show a significant deficit in the total moment below that expected for full saturation of all the Mn spins. These results suggest that the spin state of the non-ferromagnetic Mn spins is energetically well separated from the ferromagnetism of the bulk of the spins. We have also studied the coercive field ($H_c$) as a function of temperature and Mn concentration, finding that $H_c$ decreases with increasing Mn concentration as predicted theoretically.

[*] E-mail: schiffer@phys.psu.edu

The magnetic properties of diluted magnetic semiconductors (DMS) have been of interest for decades, with extensive efforts focusing on the paramagnetic, spin glass, and antiferromagnetic behavior in materials such as $Cd_{a-x}Mn_xTe$ derived from the II-VI semiconductors.[1,2] Parallel efforts have also examined hole-mediated ferromagnetism at low temperatures in IV-VI DMS alloys such as PbSnMnTe.[3] Interest in DMS materials has been recently rekindled by the realization of hole-mediated ferromagnetism in systems derived from the technologically significant III-V semiconductor GaAs.[4] Unlike the II-VI materials wherein Mn solubility poses few limitations, Mn can be incorporated in GaAs only up to concentrations of order 10%. Surprisingly, the resultant ferromagnetism is fairly robust, with ferromagnetic transition temperatures ($T_c$)[5] of up to 150K recently reported[6] and the clear formation of micron-sized domains.[7] Despite the importance of $Ga_{1-x}Mn_xAs$ as a model system, the fundamental physics underlying its magnetic properties remains the subject of considerable discussion.[8,9,10,11,12,13,14]

The $Mn^{2+}$ ions in $Ga_{1-x}Mn_xAs$ act as both a p-type dopant as well as a magnetic site in the system,[15,16,17] and the resultant holes play a significant role in the ferromagnetism of these materials. Annealing studies have also shown disorder to have a significant effect on the magnetic properties of these materials,[18,19] and recent work has suggested that the annealing may be changing the distribution of Mn interstitials that couple antiferromagnetically with Mn on the lattice sites.[20] One especially puzzling property of $Ga_{1-x}Mn_xAs$ is that the measured ferromagnetic moment[4,19,21,22,23] falls well below the expected value of 5 $\mu_B$ per Mn ion. This magnetization deficit has been attributed theoretically to glass-like freezing of spins induced by disorder[24,25,26,] or noncollinear ferromagnetism.[27] In either case, one would expect the magnetization to be increased to full saturation by relatively modest magnetic fields. However, credible measurements of the field dependence of the magnetization are hampered by the diamagnetic background of the substrates. Indirect measurements of the field dependence of the moment, such as Hall effect studies, do not provide an absolute calibration to compare the total moment to the theoretical value. In this paper, we report direct measurements of the magnetization in a series of $Ga_{1-x}Mn_xAs$ samples in magnetic fields up to 3 tesla. Contrary to earlier reports,[4] we find that -- after careful subtraction of the diamagnetic background and with careful determination of the Mn concentration -- no significant moment is recovered up to that field, setting a lower limit on the energy scale of the Mn-Mn interactions that separate the non-participating spins from the ferromagnetic state. We also have measured the coercive field ($H_c$) in this series of samples, studying the dependence of $H_c$ on temperature, Mn concentration, and post-growth annealing.

We studied a series of ferromagnetic $Ga_{1-x}Mn_xAs$ samples with $0.0135 < x < 0.083$, all of which were grown in a continuous series of increasing Mn content (except for x = 0.083, which was grown under similar conditions in a previous run). The samples for this study have been grown on (100) semi-insulating, expired GaAs substrates under conditions that have been described previously.[19] The epilayers are $123 \pm 2$ nm thick and grown on a buffer structure consisting of a standard (high temperature grown) 100 nm GaAs epilayer followed by a 25 nm low temperature grown GaAs epilayer which leads to compressive strain and hence the easy orientation of the magnetization being in-plane. X-ray diffraction measurements indicated high sample quality for the full range of Mn doping. Data are shown for either as-grown samples or for samples that were annealed at 250°C in a 99.999% purity flowing nitrogen atmosphere for 90 minutes (which yields the highest $T_c$ at that annealing temperature for a given sample thickness). Magnetization was measured in plane with a commercial superconducting quantum interference device magnetometer. X-ray diffraction and magnetization data taken to $T > 320$ K show no evidence of MnAs precipitates, although we cannot exclude the possibility of nanoscale MnAs clusters which would be superparamagnetic near the bulk MnAs $T_c$. The exact Mn concentrations were determined by electron microprobe analysis (EMPA) that is described in detail elsewhere.[19]

We measured the magnetization of annealed field-cooled samples as a function of temperature from T = 5 K to T = 200 K in magnetic fields up to H = 3 T as shown in Figure 1a. The magnetization of unannealed samples, which is somewhat lower than that of the annealed samples at low fields,[20,21] has a more complicated temperature dependence and was not studied in detail for this paper. The diamagnetic substrate contribution to this magnetization is only weakly temperature dependent except for an upturn at the lowest temperature associated with paramagnetic impurities. Thus, at temperatures well above $T_c$ where the Mn ions are all paramagnetic, the total moment of the sample is dominated by the contribution of the substrate. We can therefore subtract the substrate magnetization by normalizing the independently measured temperature dependent moment of the substrate to that of our samples at $T = 200$ K, and then subtracting this normalized background from the temperature dependent data of the sort shown in figure 1a. By repeating this procedure at different magnetic fields, we can obtain the low temperature magnetization of the $Ga_{1-x}Mn_xAs$ epilayers as a function of the field for different Mn concentrations (figure 2). It is clear from this graph that the moment per Mn atom decreases with increasing Mn concentration as noted previously.[23] Although the uncertainty at higher fields is too large to determine the magnitude of the small field dependence of the magnetization, it is clear from these data that no significant moment is recovered up to magnetic fields of this scale. If we assume the noncontributing spins are bound in antiferromagnetic pairs of interstitial Mn – substitutional Mn atoms as has been previously reported,[20] we can thus take the Zeeman energy (~ 15 K) of the $S = 5/2$ Mn spins in 3 tesla to set a lower limit on the energy scale of that coupling.

The field dependence of the magnetization is also expressed in the coercive field, an essential parameter in any potential device incorporating $Ga_{1-x}Mn_xAs$. While $H_c$ in $Ga_{1-x}Mn_xAs$ has not been the subject of detailed experimental study, both the magnetic anisotropy and the coercive field have been examined theoretically.[26,28] We have measured $H_c$ throughout the range of Mn concentrations for both as-grown and annealed samples. Typical hysteresis loops are shown in figure 1b for an annealed sample at $x = 0.0135$. Although resistance measurements have indicated in-plane anisotropy,[29] hysteresis loops for samples rotated 90° in-plane show no significant difference in their coercive fields ($H_c$). As demonstrated in figure 3, $H_c$ decreases monotonically with increasing temperature and also decreases with annealing (at least for larger Mn concentrations). The reduction of $H_c$ in annealed samples presumably results from changes in the defect structure that have been shown to be non-trivial and to couple strongly to the magnetic properties.[18,19,23] If annealing removed or weakened domain wall pinning sites, domains could expand more easily and explain the reverse in magnetization at smaller fields.

Perhaps most interesting, we find that $H_c$ is strongly reduced with increasing Mn concentration, as shown in figure 4. While the data show considerable scatter, reflecting the sensitivity of $H_c$ to details of sample preparation, this finding does corroborate mean-field theoretical predictions,[26] assuming the carrier concentration increases with Mn concentration (given by recent experimental results from Raman scattering[30] on these samples[31]).

Although there is no clear understanding yet of the field dependence of ferromagnetism in $Ga_{1-x}Mn_xAs$, our data as summarized in table 1.1 demonstrate that there is rich behavior that can be connected with recent theoretical efforts. While we have begun to probe the field-dependent phenomena, further characterization will be needed in order to control the material properties and lay the groundwork for future device applications.

N.S., S.H.C., and K.C.K were supported by ONR N00014-99-1-0071 and -0716, and DARPA/ONR N00014-99-1093. S.J.P., R.M., R.F.W., and P.S. were supported by DARPA N00014-00-1-0951 and NSF DMR 01-01318.

## Captions

Figure 1. a) Magnetic moment as a function of temperature for field cooled annealed samples at H = 0.005 T (open circle), 0.05 T (closed square), 0.5 T (closed triangle), 1 T (open triangle), and 3 T (closed circle) for x = 0.028. b) Hysteresis loops shown at different temperatures. From the widest loop to the narrowest (no loop) the temperatures are $T$ = 5 K, 10 K, 15 K, 25 K, 35 K, and 50 K for x = 0.0135. All measurements were performed with the field in-plane.

Figure 2. Magnetization of annealed $Ga_{1-x}Mn_xAs$ samples as a function of magnetic field at $T$ = 10.8 K. Data are shown subtraction of the magnetization of the substrate (as discussed in the text) and with the applied field in-plane.

Figure 3. The coercive field of $Ga_{1-x}Mn_xAs$ epilayers shown as a function of temperature for as-grown (open) and annealed (closed) samples. The data are taken from hysteresis loops similar to those shown in figure 1b.

Figure 4. The coercive field of $Ga_{1-x}Mn_xAs$ epilayers at T =5 K shown as a function of carrier concentration determined by Raman scattering for as-grown (open) and annealed (closed) samples.

Table 1. Summary of important sample parameters in this study.

| Annealed | | | | |
|---|---|---|---|---|
| Mn (%) ± 0.2 | $T_c$ (K) ± 2K | $H_c$ @ 5K (Oe) | $M_s$ ($\mu_B$/Mn) | P ($10^{19}$) |
| 1.35 | 42 | 114 ± 3 | 4.6 ± 0.15 | 3.8 ± 0.4 |
| 1.72 | 50 | 72 ± 3 | 4.6 ± 0.15 | 6.5 ± 0.6 |
| 2.78 | 70 | 52 ± 3 | 3.98 ± 0.10 | 16.0 ± 2 |
| 3.34 | 80 | 24 ± 3 | 4.13 ± 0.10 | 27.1 ± 3 |
| 5.54 | 110 | 4.8 ± 1 | 2.98 ± 0.05 | 63 ± 6 |
| 5.97 | 110 | 13 ± 3 | 2.63 ± 0.05 | 55 ± 6 |
| 8.3 | 110 | 27 ± 3 | 1.74 ± 0.05 | 71 ± 7 |

| As-grown | | | |
|---|---|---|---|
| Mn (%) ± 0.02 | $T_c$ (K) ± 2K | $H_c$ @ 5K (Oe) | P ($10^{19}$) |
| 1.35 | 42 | 117 ± 3 | 3.6 ± 0.4 |
| 3.34 | 60 | 40 ± 3 | |
| 5.54 | 80 | 21 ± 3 | 23.2 ± 2 |

Table 1. Potashnik et al.

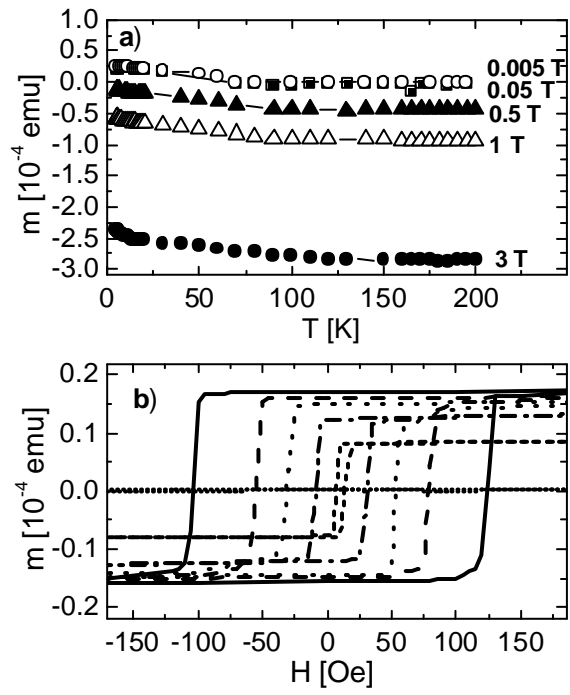

Fig 1. Potashnik et al.

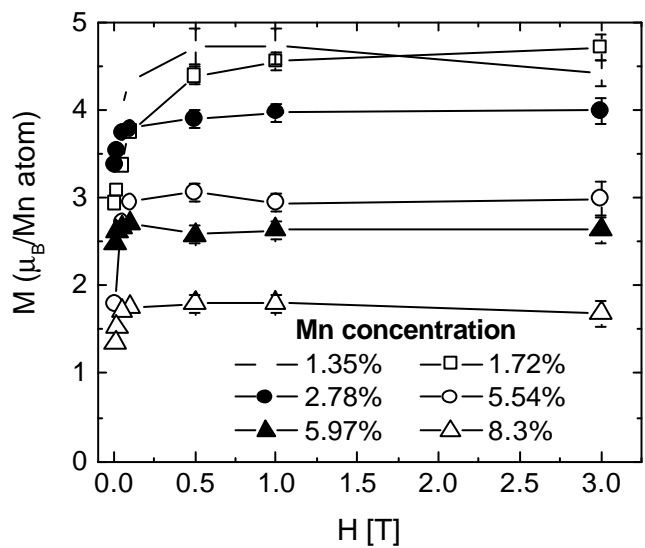

Figure 2. Potashnik et al.

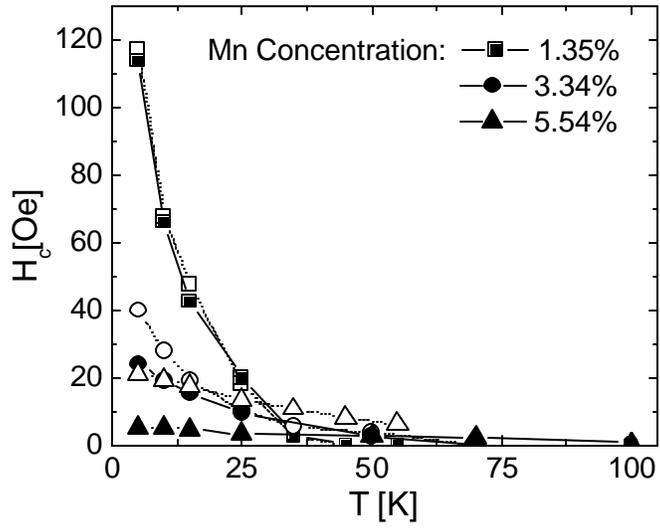

Fig 3. Potashnik et al.

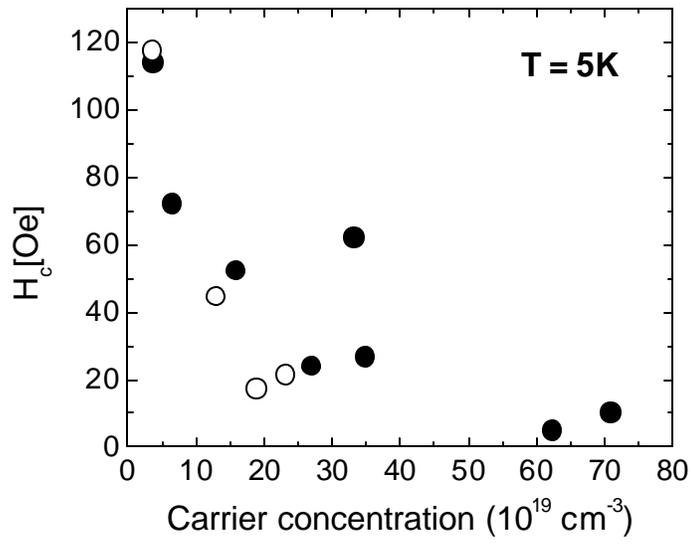

Figure 4. Potashnik et al.